\documentclass[aps,prb,twocolumn,showpacs,superscriptaddress,floatfix,nofootinbib]{revtex4-1}

\usepackage{xcolor} \usepackage{graphicx} \usepackage{textcomp}
\usepackage{amsmath} \usepackage{epsfig} \usepackage{helvet}
\usepackage{amssymb}

 \newcommand{\be}{\begin{equation}}
  \newcommand{\ee}{\end{equation}} \newcommand{\bea}{\begin{eqnarray}}
  \newcommand{\eea}{\end{eqnarray}}

\newcommand{\tr}{\mbox{tr}} \newcommand{\nn}{\nonumber}

\newcommand{\ket}[1]{\mbox{$| #1 \rangle$}}
\newcommand{\braket}[2]{\mbox{$\langle #1 | #2 \rangle$}}

\DeclareGraphicsExtensions{.pdf,.png,.jpg}

     \def\tr{ \mbox{tr}} 

\begin{document}
 
\title{Dynamical windows for real-time evolution with matrix product
  states}

\author{Ho N. Phien} \affiliation{Centre for Engineered Quantum
  Systems, School of Mathematics and Physics, University of
  Queensland, Brisbane 4072, Australia}

\author{Guifr\'e Vidal} \affiliation{Perimeter Institute for
  Theoretical Physics, Waterloo, Ontario, N2L 2Y5, Canada}

\author{Ian P. McCulloch} \affiliation{Centre for Engineered Quantum
  Systems, School of Mathematics and Physics, University of
  Queensland, Brisbane 4072, Australia}

\date{\today}

\begin{abstract}
  We propose the use of a dynamical window to investigate the
  real-time evolution of quantum many-body systems in a
  one-dimensional lattice. In a recent paper [H. Phien et al,
  arxiv:????.????], we introduced infinite boundary conditions (IBC)
  in order to investigate real-time evolution of an infinite system
  under a local perturbation. This was accomplished by restricting the
  update of the tensors in the matrix product state to a finite
  window, with left and right boundaries held at fixed positions. Here
  we consider instead the use of a dynamical window, namely a window
  where the positions of left and right boundaries are allowed to
  change in time. In this way, all simulation efforts can be devoted
  to the space-time region of interest, which leads to a remarkable
  reduction in computational costs. For illustrative purposes, we
  consider two applications in the context of the spin-1
  antiferromagnetic Heisenberg model in an infinite spin chain: one is
  an expanding window, with boundaries that are adjusted to capture
  the expansion in time of a local perturbation of the system; the
  other is a moving window of fixed size, where the position of the
  window follows the front of a propagating wave.
\end{abstract}
\pacs{03.67.-a, 03.65.Ud, 02.70.-c, 05.30.Fk}

\maketitle

\tableofcontents
\section{Introduction}
Ever since the density-matrix renormalization group method
(DMRG)\cite{White1, White2} was invented in 1992 by Steven R. White,
it has opened new trends for studying numerically strongly correlation
effects of quantum systems in one dimension. By now, it is well
established as a powerful method in producing numerically exact
results for ground state wavefunctions and expectation values of
one-dimensional quantum systems. In addition, the DMRG is not
constrained itself within a small regime for investigating static
properties but it has been extended to study dynamical
properties\cite{Hallberg1,Kuhner1,Jeck1,Cazalilla1} as well as quantum
systems at finite temperature\cite{Wang1,Sirker1}.

The connection was
not immediately made that the wavefunction produced by DMRG can be
realized as a variational calculation in the space of matrix product
states (MPS)\cite{Ostlund1, Fannes1,Perez1}. As it is much simpler and
easier, generally people prefer to implement the DMRG in terms of MPS.
Furthermore, together with MPS the tensor network (TN) ansatzs have
been attracting much interest from computational physicists.
Algorithms have developed based on MPS to simulate both static and
dynamical properties of 1D quantum systems. One of the most successful
algorithms in MPS formalism is the time-evolving block decimation
(TEBD) algorithm\cite{Vidal1,Vidal2}, which has an equivalent DMRG
formulation\cite{White3,Daley1}.  This algorithm can be used for
ground state calculations, although it is not as efficient as
variational minimization algorithms such as DMRG. However TEBD comes
into its own for real-time evolution.  More recently, a new algorithm
called the time-dependent variational principle (TDVP)\cite{Haegeman}
has also been introduced.

There are many interesting problems that involve dynamics of a small
section embedded in an infinite lattice. For example, consider the
real-time evolution of an infinite quantum spin chain after one site
in the middle of the chain has been locally perturbed by some spin
excitation, e.g. a local $S^{+}$ operator. The ground state is now modified and become
a superposition of excited states. Therefore the infinite MPS (iMPS)
cannot be represented in a translationally invariant form anymore.
This is a huge hurdle to investigate real-time evolution of the system
in thermodynamic limit, as in principle one must use an
infinite set of different tensors in the iMPS to describe the wave
function of the state. This makes the simulation an impossible
task. However, because the perturbation is local, one can avoid this
difficulty and can still understand the dynamical properties of a
system in thermodynamic limit, as long as the range of effect is
finite (or approximately so).  The conventional way to solve this problem is
to use a large, but finite lattice chosen to be large enough that the
boundaries are far enough away to not cause problems for the
calculation. As the wavefunction evolves in time, 
this local perturbation represented by a
wave-packet will spread throughout the system at the group velocity (for ballistic transport)
or slowly spread out through the system (in the diffusive case).
Thus, to obtain the long-time evolution of a system in this way the lattice
size must be fairly large, firstly to avoid Friedel oscillations from the
boundaries affecting the ground state, and also for the time evolution 
the simulation typically needs to stop once the wave front approaches
the boundary.  In a previous paper\cite{PhienPrevious} we have shown
how the computational cost can be substantially improved by using a
much smaller finite system with infinite boundary conditions (IBC).
This has two advantages, firstly there is no hard boundary in the
system, so no Friedel oscillations, 
and away from the perturbation the system is asymptotically
translationally invariant; and secondly, since the `boundaries' of the
finite system represent an effective semi-infinite chain rather than a
hard wall, there is no problem to allow the perturbation to propagate
beyond the finite region as long as the wavefunction doesn't
move too far outside the effective Hilbert space of the semi-infinite chain. 
This is a big advantage over traditional
finite-size calculations.  To achieve this, we divide the whole spin
chain into three parts where the middle part contains the
perturbation, called the \emph{window}, and other two parts on the
left and right of its which are not affected by the perturbation. The
boundaries of the window are represented by an effective Hilbert space
for the wave function on a semi-infinite strip.

In this paper, we further improve the computational efficiency of the
IBC technique by focusing on how the wavefront propagates in time.
Efficiencies can be obtained by introducing dynamical window
techniques, namely expanding and moving the window throughout the
calculation.  Specifically, we will keep track of the wavefront and
decide to expand or move the window such that the physically relevant
section of the system is well represented.  An important point is that
in our scheme the section of the system outside of the window,
represented by an effective Hilbert space, can evolve as well so that
the window size can be quite small, containing only on the region of
interest, without affecting the accuracy too much.  An Results are
presented for a the evolution of a local perturbation in the spin-1
antiferromagnetic (AFM) Heisenberg model.

\section{Dynamical window techniques}
Let us consider an infinite spin chain where the ground state is
represented by a translationally invariant infinite matrix product
state (iMPS) with a one-site unit cell, 
\bea \ket{\Psi}
=\sum_{s_{i}}\ldots\lambda\Gamma^{s_{i}}\lambda\Gamma^{s_{i+1}}\ldots\ket{\bf{s}},
\label{eq1}
\eea 
where $\ket{\bf{s}}=\ket{\ldots s_{i},~s_{i+1}\ldots}$ is the
basis in Hilbert space $\mathcal{H}\in\mathbb{C}^{\otimes d}$ of the
system ($d$ is the dimension of local Hilbert space at each lattice
site). This iMPS can be always written in the mixed canonical form as
following 
\bea \ket{\Psi} = \sum_{s}\ldots A^{s_{i-1}}A^{s_{i}}\lambda
B^{s_{i+1}}B^{s_{i+2}}\ldots\ket{\bf{s}},
\label{eq2}
\eea 
where tensors $A$ and $B$ satisfy the canonical form constraints
\bea \sum_{s_{i}}{A^{s_{i}}}^{\dagger}A^{s_{i}}
&=&\sum_{s_{i}}{\Gamma^{s_{i}}}^{\dagger}\rho^{R}\Gamma^{s_{i}}=\mathbb{I},
\label{constraint1}\\
\sum_{s_{i}}B^{s_{i}}{B^{s_{i}}}^{\dagger}
&=&\sum_{s_{i}}\Gamma^{s_{i}}\rho^{L}{\Gamma^{s_{i}}}^{\dagger}=\mathbb{I},
\label{constraint2}
\eea 
where $\mathbb{I}$ is identity matrix, $\rho^{L}$ and $\rho^{R}$
are the left and right reduced density matrices, respectively.

After perturbing the ground state, the infinite spin chain can be
described effectively by the finite MPS as, 
\bea \ket{\tilde{\Psi}} =
\sum_{\{s_{i}\}} L^{\alpha} A_1^{s_{1}}\lambda A_2^{s_{2}}\ldots
A_N^{s_{N}} R^{\beta} \: \ket{\alpha, \mathbf{\tilde{s}}, \beta},
\label{eq3}
\eea 
where $\ket{\bf{\tilde{s}}}=\ket{s_{1}, s_{2}, \ldots s_{N}}$,
$A_1^{s_{1}}$ fulfills the condition in Eq.~\ref{constraint1} and all
the other tensors on the right of $\lambda$ matrix satisfy the right
canonical constraint in Eq.~\ref{constraint2}. Calculation for
corresponding effective Hamiltonian of this system was described in
Ref.~\onlinecite{PhienPrevious} as was the scheme for evolving the
system in time using TEBD algorithm. Here, we reuse these techniques
combining with the dynamical window technique to investigate the
dynamical properties of the system. Dynamically changing the window
size involves two basic steps, contraction of the window and expansion
of the window. We now describe the technical steps involved in each
case.

\subsection{Window Expansion}

\begin{figure}[ht]
  \centering
  \includegraphics[scale = 0.8]{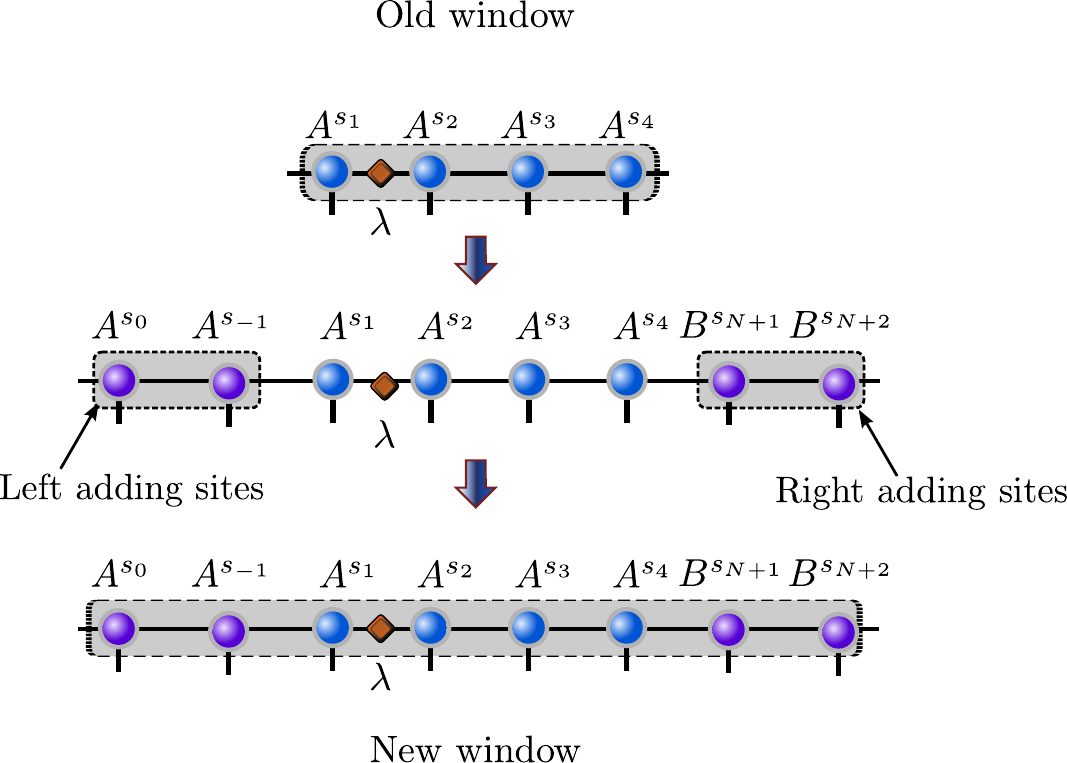}
  \caption{(Color online): Diagrammatical representation of adding the
    sites into both sides of the window for the window expansion
    step.}
  \label{fig:ExpandWMPS}
\end{figure}

Expanding the window incorporates more degrees of freedom into the variational
wavefunction, and this is an operation that one will typically want to do in order
to follow the propagation of a perturbation as it travels through the lattice.
This is achieved by incorporating some
sites from the translationally-invariant semi-infinite chain
into the finite-size window. The window can be expanded on the left
and the right hand side separately, for example to follow the wavefront
of a symmetrically-expanding local perturbation we use the scheme in Fig.~\ref{fig:ExpandWMPS}.
The basic operation on the the MPS, 
from Eq.~\ref{eq3} becomes, in the case of a two-site unit cell expanding on the left-hand
side,
\bea
\ket{\tilde{\Psi}} = \sum_{s_{i}}L^{\alpha}A_{-1}^{s_{-1}}A_0^{s_{0}}A_1^{s_{1}}\lambda
A_2^{s_{2}}\ldots A_N^{s_{N}} \: \ket{\alpha \mathbf{\tilde{s}'} \beta},\\
\eea 
where the initial values of the tensors $A_{-1}$ and $A_0$ are simply given by
the translationally-invariant matrices $A$ of the ground state Eq.~\ref{eq2}. Note that
the Hilbert space $\ket{\alpha}$ at the left edge of the window is unchanged.
Therefore the block operators acting on this space are also unchanged, although
care needs to be taken that the energy of the system is correctly taken into account.
In calculating the effective Hamiltonian\cite{PhienPrevious}, the energy per site
of the infinite system appears as a separate term which is easily removed. Thus,
in order for the total energy of the system to remain constant as the window is
expanded, it is convenient to subtract this energy off the Hamiltonian for the finite
window as well. This amounts to subtracting the ground state energy per site off the
window Hamiltonian for each site added to the window.

\subsection{Window Contraction}

\begin{figure}[ht]
  \centering
  \includegraphics[scale = 0.8]{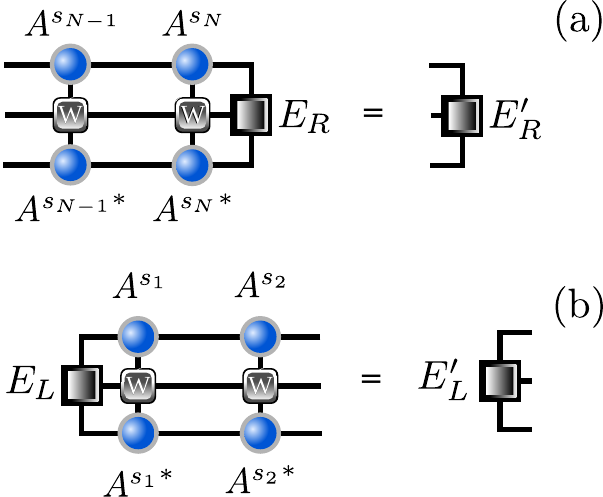}
  \caption{(Color online): Updating the effective Hamiltonian for
    contracting the window size. (a). Right update is performed when 
    incorporating sites from the right-hand edge of the window into the right boundary.
    (b). Left update for incorporating sites from the left-hand edge of the window
    into the left boundary.}
  \label{fig:Eff_Oper1}
\end{figure}

The second operation that we can perform on the window is to contract the size of it,
by absorbing some sites of the window into the boundary tensor. To achieve
this, we contract over those sites to obtain a new set of block operators and effective
Hamiltonian that will now describe a semi-infinite chain plus some number of
additional (not translationally invariant) sites. This procedure is 
implemented as shown in Fig.~\ref{fig:Eff_Oper1}.
The components of the block operators become, for example on the right-hand side,
\bea E'_{R} =
\sum_{s_{N}}\sum_{s_{N-1}}{A^{s_{N-1}}}^{\dagger}{A^{s_{N}}}^{\dagger}
E_{R}{A^{s_{N}}}{A^{s_{N-1}}}\nn\\
\braket{s_{N}}{W|s_{N}}\braket{s_{N-1}}{W|s_{N-1}},
\label{eq12}
\eea 
where $W$ is the matrix product operator (MPO) of the Hamiltonian
of the system. The tensors $A^{s_{N}}$ and $A^{s_{N-1}}$ satisfy the
right canonical form constraint in Eq.~\ref{constraint2} and in the
Fig.\ref{fig:MPSMovingW1}, they are $A^{s_{4}}$ and $A^{s_{3}}$,
respectively.

Again, one needs to take care that the total energy of the system
is unchanged in this procedure, which is easily effected by
adding a constant equal to the ground state energy per site to the
Hamiltonian of the window.

\subsection{Moving window criteria}
Now we will explain the criteria when to expand or move the window.
There are many ways to do this, for example, from the result of fixed window, 
we can look at how the wavefront propagates in time and 
determine the maximum velocity of excitations.
However, it is much more convenient to use another
criteria that relies on the fidelity of the tensors which include
$A^{s_{1}}$ and $\lambda$ matrices between two successive time steps at
the edge of the window. We do this by measuring how much the reduced density matrix
changes from one time-step to the next.
Assume that at time $t$ the MPS is described by Eq.~\ref{eq3} with the
reduced density matrix $\rho^{R} = \lambda^{2}$ and later time
$t+\delta t$ is represented by 
\bea 
\ket{\tilde{\Psi}'} =
\sum_{s_{i}}B^{s_{1}}\lambda' B^{s_{2}}B^{s_{3}}\ldots
B^{s_{N}}\ket{\bf{\tilde{s}}},
\eea 
where the new reduced density matrix ${\rho^{R}}' =
\lambda'^{2}$. If two tensors $A^{s_{1}}\lambda$ and
$B^{s_{1}}\lambda'$ are the same then ${\rho^{R}}'=\rho^{R}$. 
What we want to check is the fidelity of these two density matrices,
\bea
\sqrt{F}({\rho^{R}}', \rho^{R}) = \tr \sqrt{\sqrt{\rho^{R}} {\rho^{R}}' \sqrt{\rho^{R}}}
\eea
This fidelity is obtained as the sum of the singular values of
\bea
\sum_{s_{1}}{(B^{s_{1}}\lambda')}^{\dagger}A^{s_{1}}\lambda =
USV^{\dagger}. 
\eea
If this fidelity is close to 1 then 
we can conclude
that these tensors are the same and we do not need to move the window. 
Typically, we use a threshold of $1 - 10^{-4}$ in our numerical calculations.

\section{Numerical Calculations}

Now we explain in detail how to calculate the observables of the
system during the time evolution of the system with increasing window
size. The two observables that we are interested in are: local
magnetization $\langle S^{z}(x,t)\rangle$ ($x$ is the position of the
lattice site) in which we can see how the wave packet propagates in
time and un-equal time two-point correlator $\langle A(x,t)\rangle$
from which one can extract the spectral function. We calculate these
observables for two different dynamical window schemes; an
expanding window, where we increase the size of the window symmetrically 
to encompass the symmetrically expanding wavefronts, and a moving
window where we take the window to be much smaller, to test the case
where one is interested mainly in the dynamics of a small section of
a larger system. In the latter case, a significant amount of the dynamics
will occur outside of the window, so an important test of the method is
to check that the dynamics within the window remains accurately described.
In this section, we present results for the spin-1 AFM Heisenberg model. 
Simulation has been
implemented with $\chi = 200$ where $\chi$ is number of states kept in
TEBD. The time step is $\delta t = 0.05$ and we have used fourth order
Suzuki-Trotter expansion\cite{Suzu1}.

\subsection{Expanding window}

\begin{figure}[ht]
  \centering
  \includegraphics[scale = 0.5]{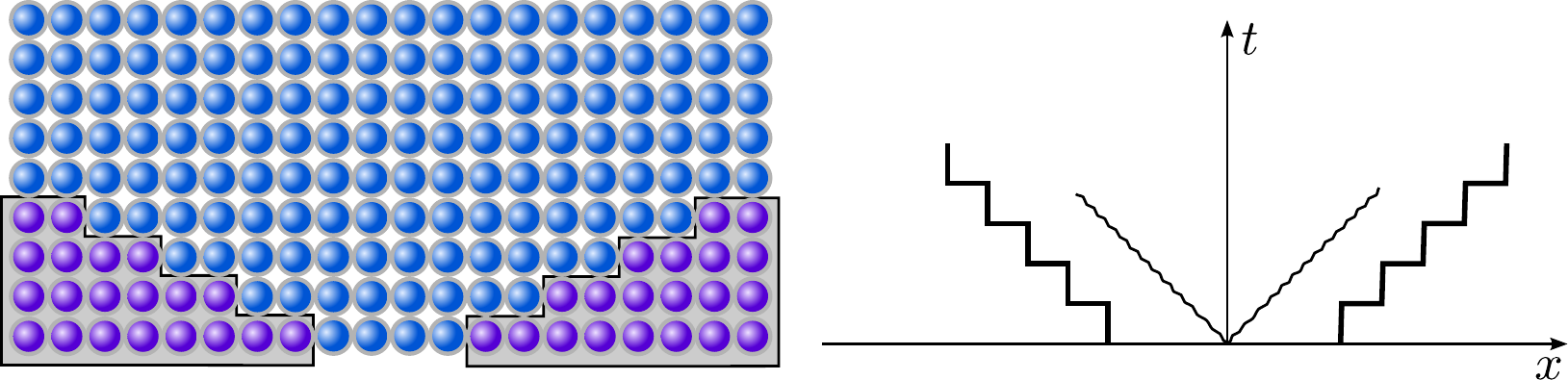}
  \caption{(Color online): Illustration of how the window is expanded
    in time and space. The balls are represented for the lattice sites
    on the left figure.  Blue balls are inside the windows and outside
    the window, translationally invariant purple balls are used. On
    the right figure, the wavefront is moving in time while the window
    is expanded along the spin chain axis. }
  \label{fig:ExpandW}
\end{figure}

We start the time evolution of the system with a small window
size $N_{e}$ in which the perturbation appears in the middle at
position $N_{e}/2$ from the left. Outside the window, the MPS
tensors are position-independent. The expanding window scheme is illustrated
in Fig.~\ref{fig:ExpandW}. The starting window contains $N_{e}=4$
lattice sites and then increases with time. We can see that when the
wavefront hits boundaries, the window moving criteria is met so we need
extend the window by adding some number of unit cells to both sides of
the window. Here we add one two-site unit cell at a time, 
however in principle, we can add as many unit
cells as we want. After adding the sites, we evolve the
system for some more time before the next expanding procedure.

\begin{figure}[ht]
  \centering
  \includegraphics[scale = 0.5]{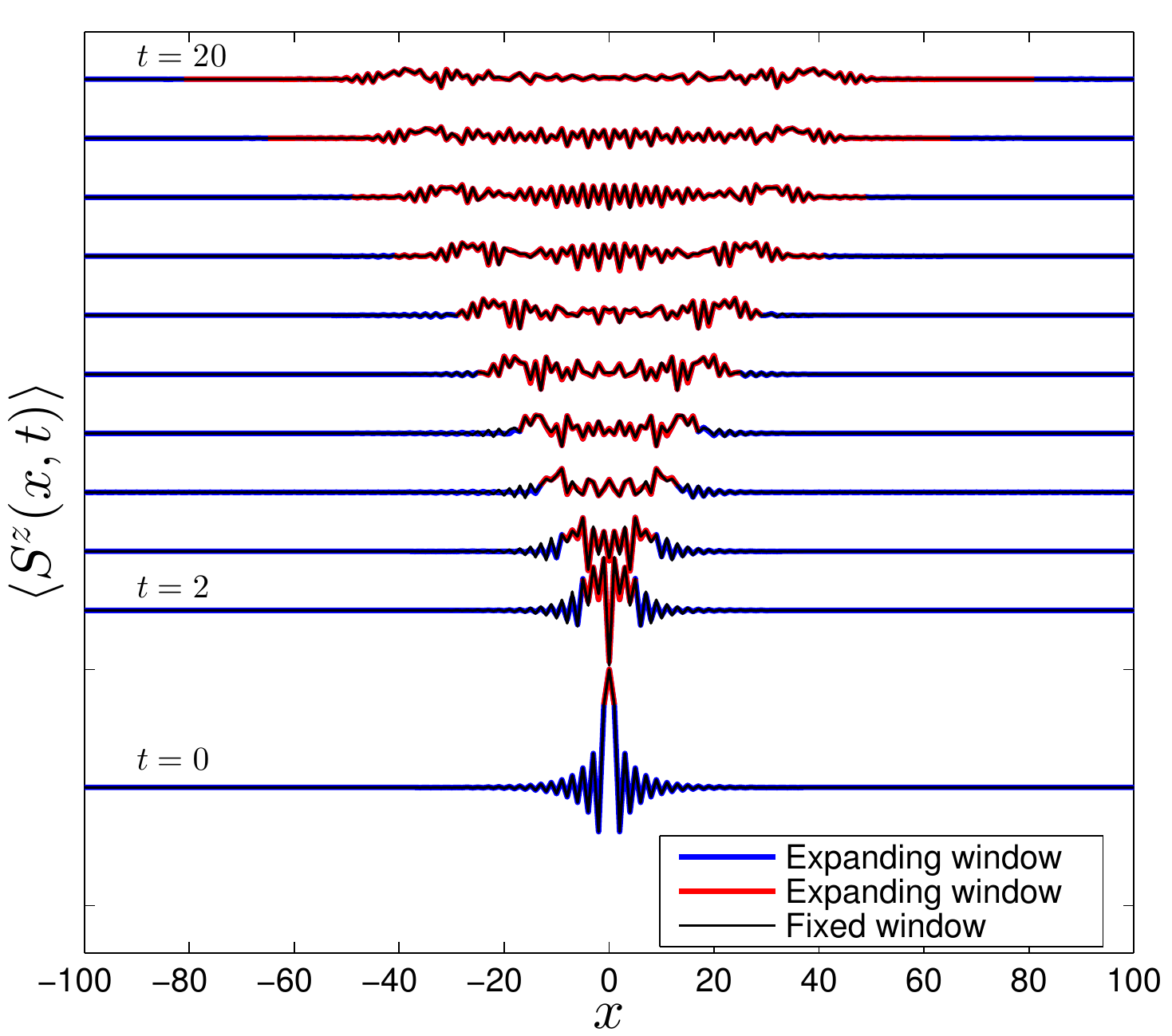}
  \caption{(Color online): Comparison of wave-packet propagating in
    time between different schemes: fixed and expanding window.}
  \label{fig:SzTimeExpandingW1}
\end{figure}

\begin{figure}[ht]
  \centering
  \includegraphics[scale = 0.5]{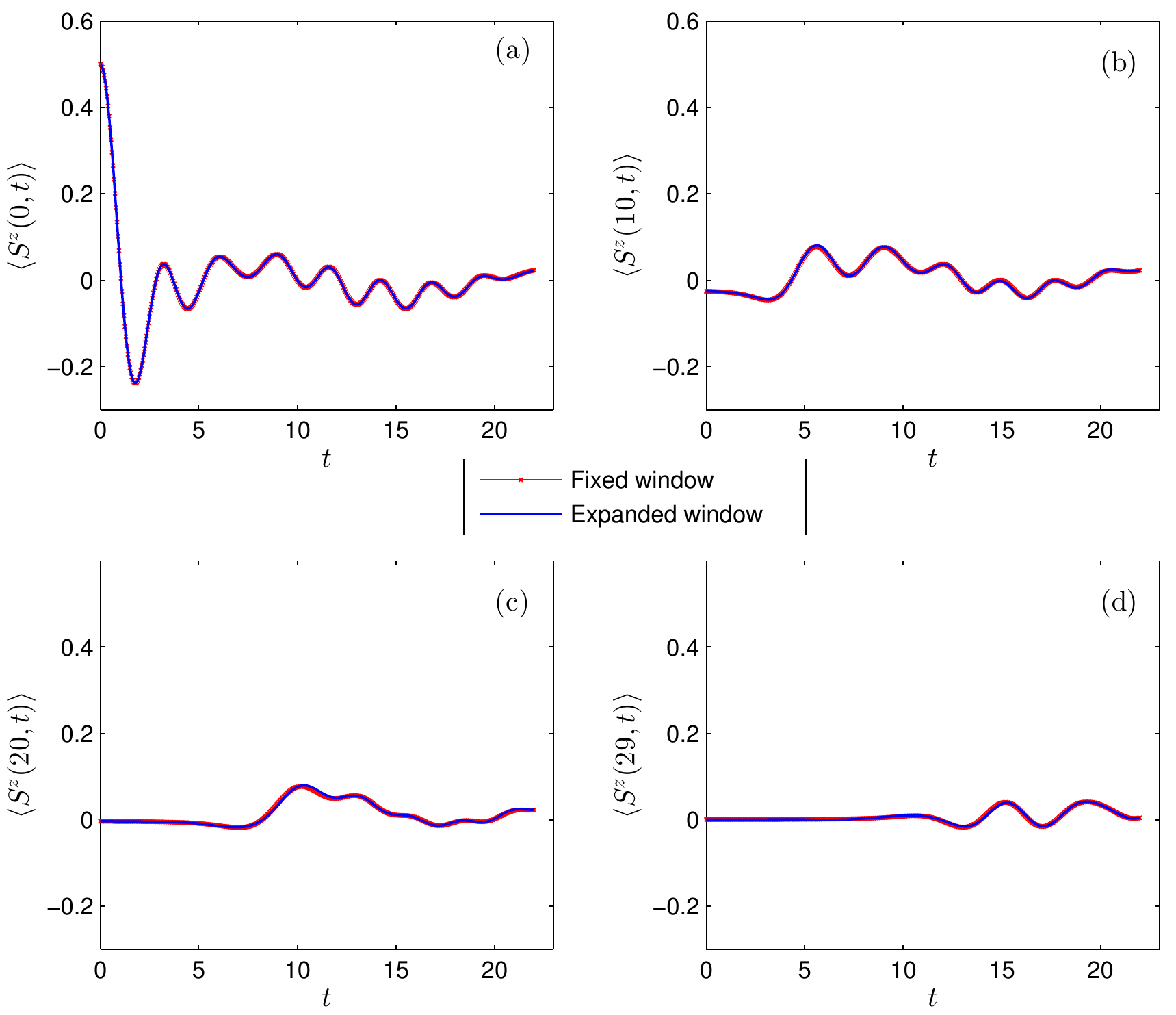}
  \caption{(Color online): Comparison of local magnetization evolving
    in time at specific positions $x_{j}$ between different schemes:
    fixed and expanding window. (a). $x_{j} = 0$; (b). $x_{j} = 10$;
    (c). $x_{j} = 20$; (d). $x_{j} = 29$. }
  \label{fig:SzTimeExpandingW2}
\end{figure}

Fig.~\ref{fig:SzTimeExpandingW1}
shows our numerical calculation of $\langle S^{z}(x,t)\rangle$. 
It also shows the result
calculated for the case of time-evolving fixed window of 
size $N_{f}=240$ for comparison (this is numerically exact for these purposes).
It can be seen that the results obtained from the
different methods are almost the same although the size of starting
window is small and increasing slowly in time. Of course, the
expanding window calculation is much faster because the computation
time is essentially linear in the size of the finite window.

\begin{figure}[ht]
  \centering
  \includegraphics[scale = 0.5]{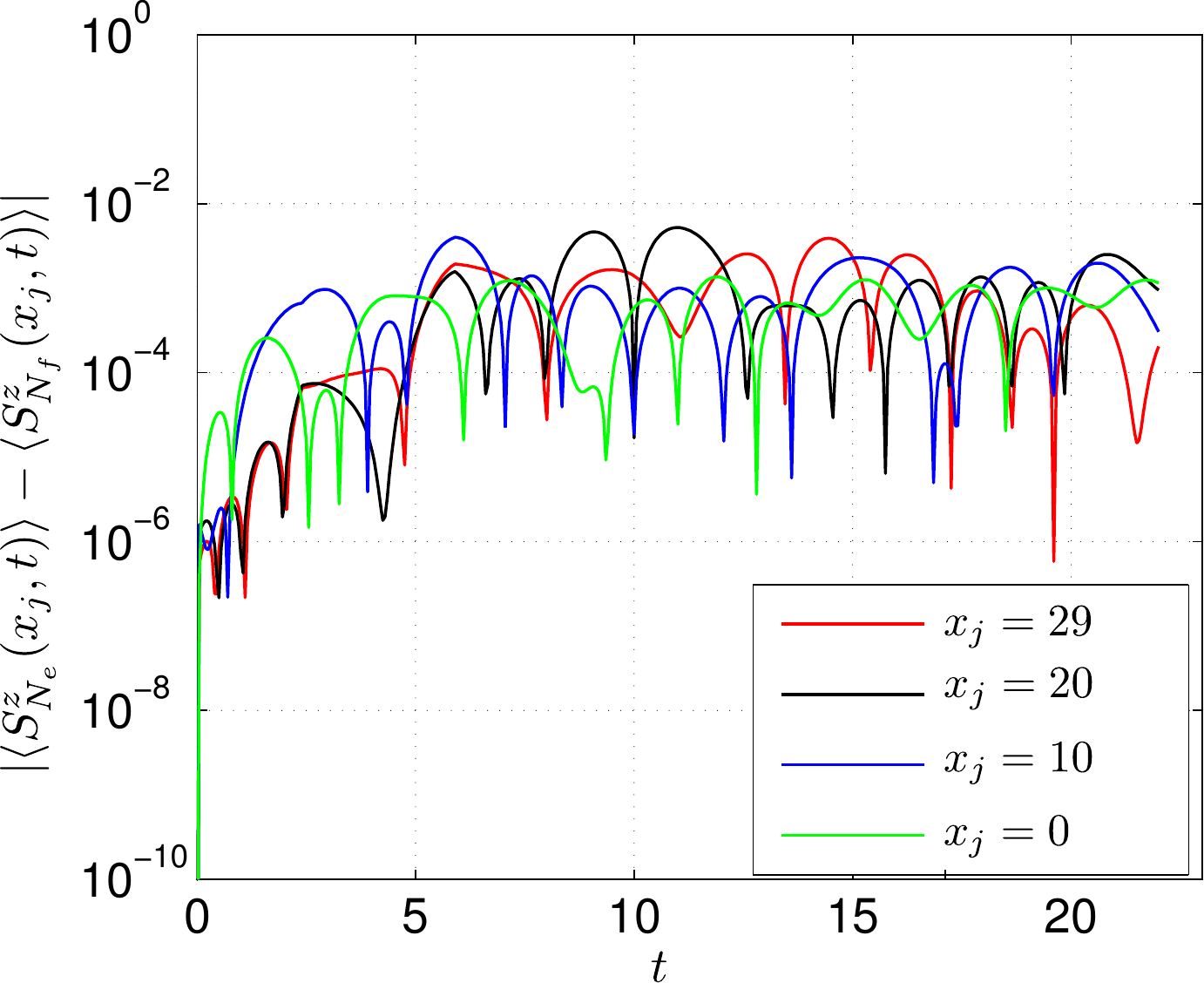}
  \caption{(Color online): Difference in local magnetizations evolving
    in time at specific positions $x_{j}$ between different schemes:
    fixed and expanding window. }
  \label{fig:SzTimeExpandingW3}
\end{figure}

In order to see how different the wave packets between expanded and
fixed window methods are, we plot the local magnetization $\langle
S^{z}(x_{j},t)\rangle$ at different positions $x_{j}$ corresponding to
the initial local perturbation position of equivalent lattice sites.
The results are shown in Fig.~\ref{fig:SzTimeExpandingW2}, $x_{j} =
\{0, 10, 20, 29\}$. We can see that they match very
well. For a better understanding, we plot of absolute difference
between two methods in Fig.~\ref{fig:SzTimeExpandingW3}. By now it can
bee seen that these differences increases with time. However, they all
oscillate around an acceptably small value $\sim 10^{-4}$.

\subsection{Moving window}

\begin{figure}[ht]
  \centering
  \includegraphics[scale = 0.5]{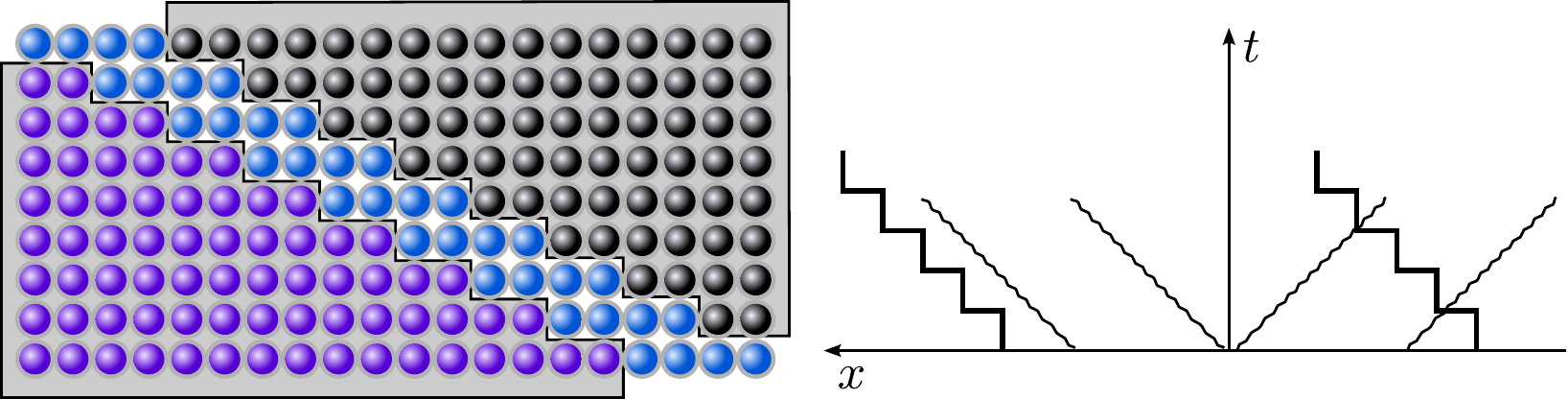}
  \caption{(Color online): Illustration of how the window is shifted
    in time and space. The balls are represented for the lattice sites
    of the left figure. Inside the windows blue balls are used. On the
    left side the window, translationally invariant purple balls are
    used. The black balls corresponds to the blue balls at the same
    positions of the chain before moving the window. Figure on the
    right shows the wavefront moving in time along the space while the
    window is shifted.}
  \label{fig:WMoving1}
\end{figure}

We now discuss another scheme of dynamical window technique, 
which combines the expansion and contraction steps to obtain
a fixed-size but moving
window. The window is now fixed in size and is shifted along the chain
as soon as the wavefront hits boundary. In principle, we can move this
window to either sides of chain. In Fig.~\ref{fig:WMoving1}, the
initial small-size window which contains $N=4$ sites and is located on
the left most side of the finite chain. As the wavefront propagates in
time, this window will move to the left of the spin chain. As a
consequence of this, on the right side of the window, the wavefront
will hit boundary after some time but on the left side the wavefront
doesn't hit the boundary. Therefore, all the information of dynamical
properties measured in the area containing blue and purple balls should
be reliable, but the wavefunction in the black region will be obtained
only in a small effective Hilbert space.

The advantage of this scheme is that it is very cheap in
computational cost as we just need to modify and update the sites
inside the window. To understand how it is implemented in terms of
tensor network, we introduce the update scheme for MPS and effective
Hamiltonian before the window is shifted to a new position. There are
two most important components in the update scheme of the real-time
evolution algorithm of the MPS with infinite boundary conditions.

\begin{figure}[ht]
  \centering
  \includegraphics[scale = 0.8]{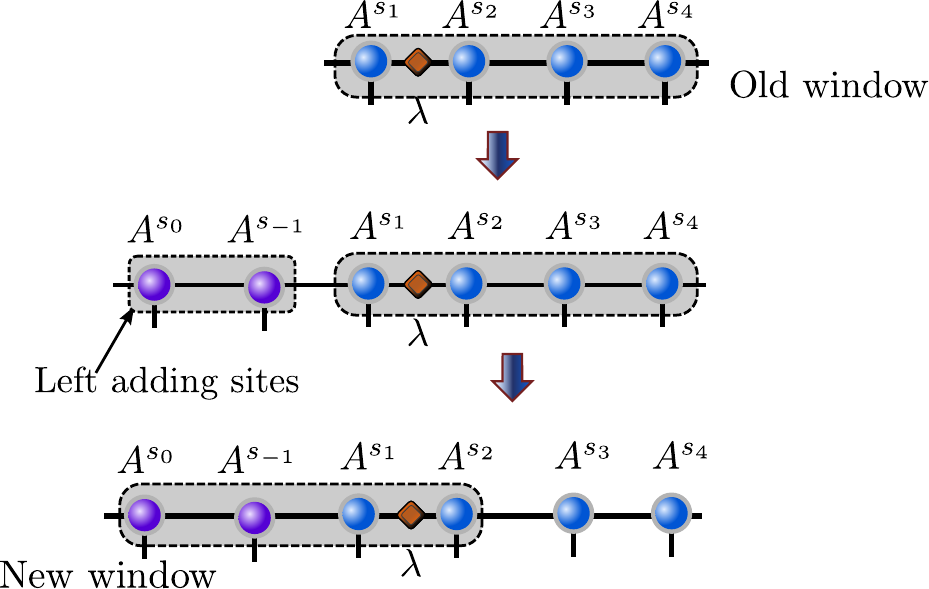}
  \caption{(Color online): Illustration of how to update the MPS when
    moving the window to the left of the spin chain.}
  \label{fig:MPSMovingW1}
\end{figure}

The procedure for updating the new window contained in the MPS is
described in Fig.~\ref{fig:MPSMovingW1}. When the moving window
criteria is met, we need to shift the window to the left by one
two-site unit cell.
Note that
the number of sites in the old window that are absorbed into the right
boundary is equal to the number of added sites to make sure that the
size of the new window is as same as the old one.

\begin{figure}[ht]
  \centering
  \includegraphics[scale = 0.5]{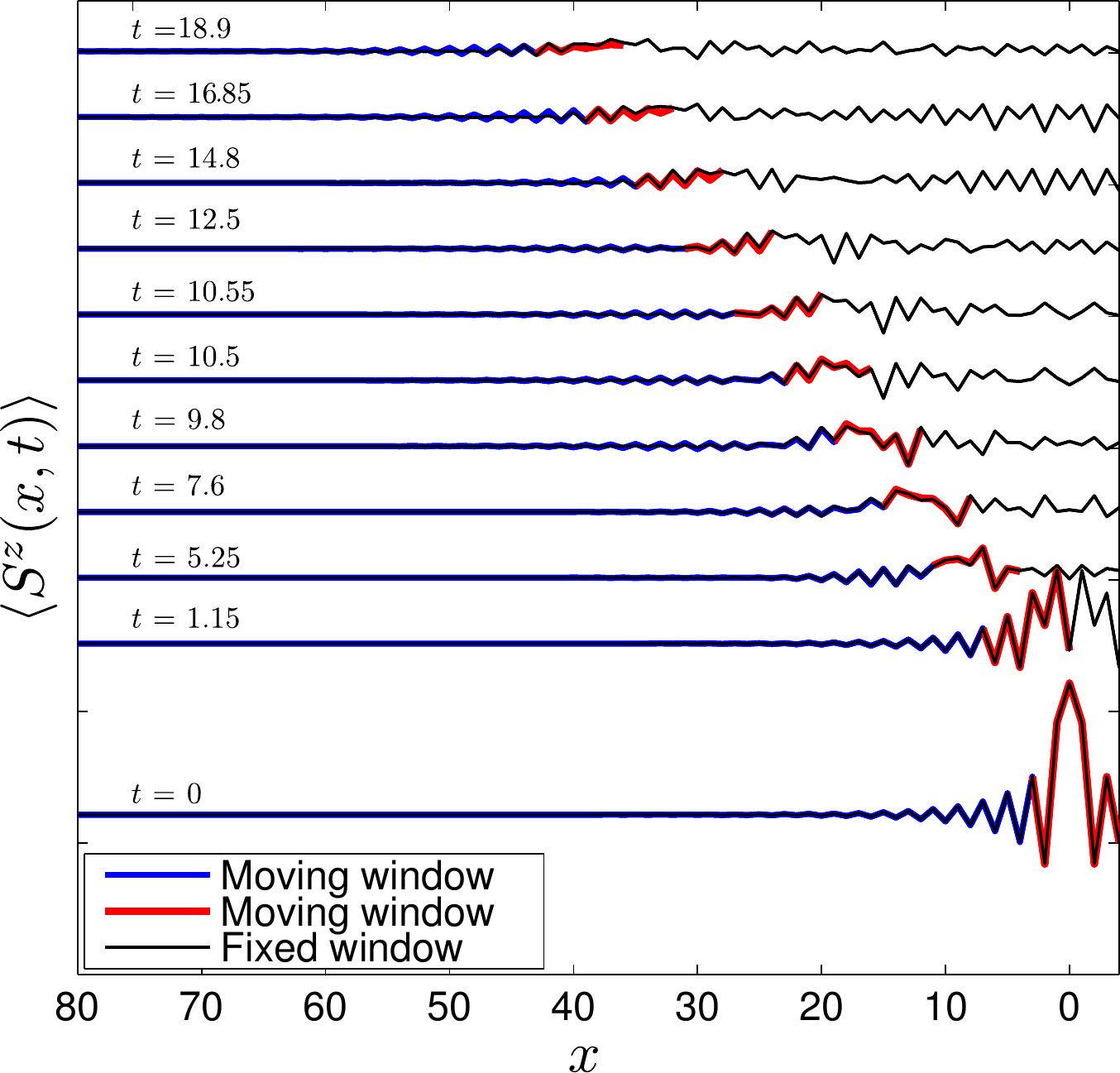}
  \caption{(Color online): Comparison of wave-packet propagating in
    time between different schemes: fixed and moved window. The black
    lines corresponds to the fixed window with size $N_{f} = 240$. The
    red and blue lines correspond to the cases of inside and outside
    of the moved window size $N_{m}=8$, respectively.}
  \label{fig:SzTimeMovingSmallW4}
\end{figure}

Again, we investigate the local magnetization $\langle
S^{z}(x,t)\rangle$ and compare with the result obtained for the fixed
window case. The comparison is plotted in
Fig.~\ref{fig:SzTimeMovingSmallW4}. In our numerical calculation
the window size of moving window
scheme is chosen to be $N_{m}=8$ compare to the size of fixed window
$N_{f} = 240$. We can see that the moved window captures well the
wavefront propagating to the left of the chain and fit nicely with the
case of fixed window. Therefore, we have obtained most of the dynamical
information of interest, with a calculation that is an order of magnitude
more efficient than choosing a large fixed window, and therefore
\emph{much} more efficient than the traditional method for performing
this calculation, namely a large finite system with open open boundary conditions.

\begin{figure}[ht]
  \centering
  \includegraphics[scale = 0.5]{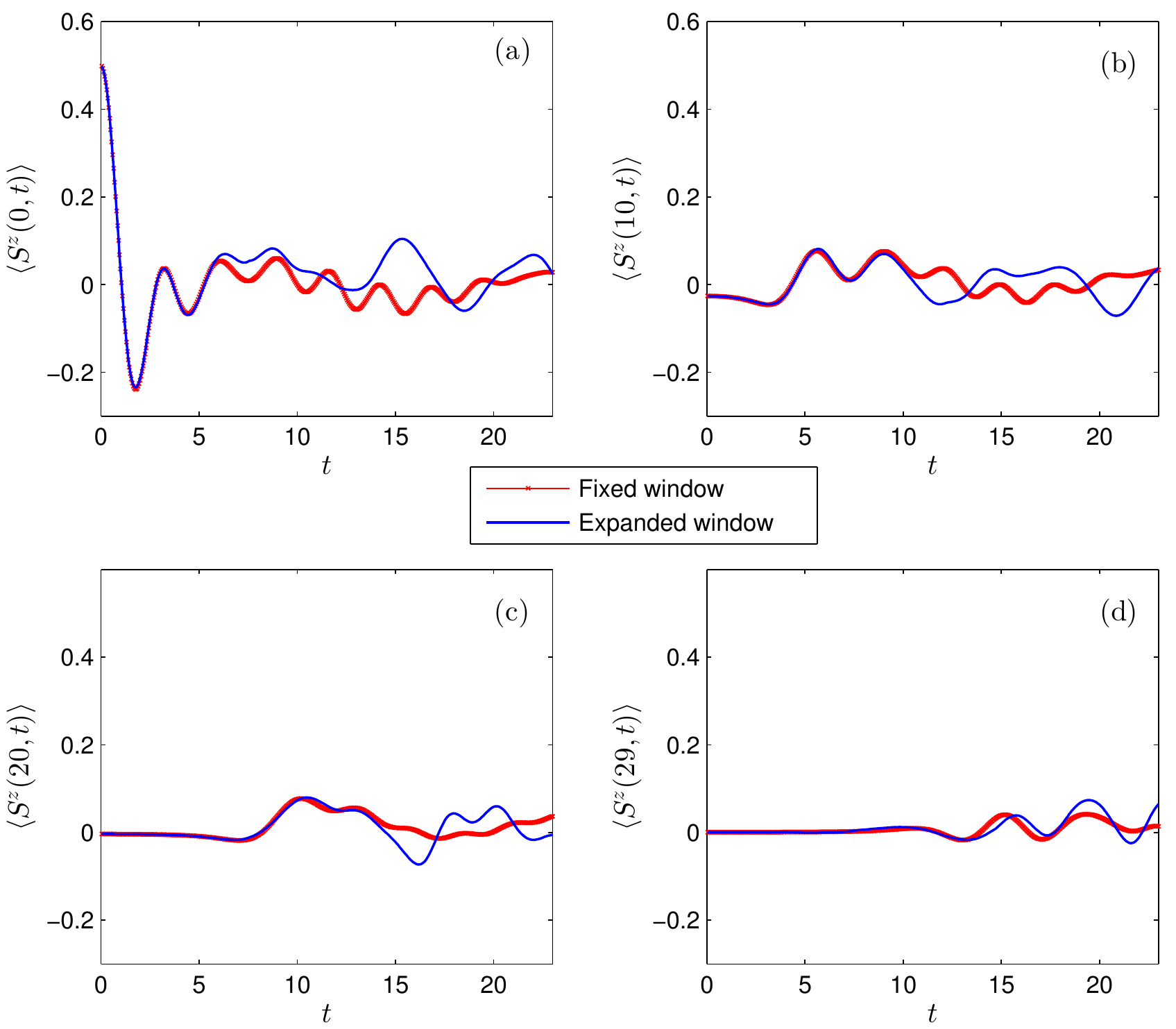}
  \caption{(Color online): Comparison of local magnetization evolving
    in time at specific positions $x_{j}$ on the left side of
    perturbation point between different schemes: fixed and moving
    window. The size of fixed window is $N_{f}=240$ and $N_{m}=8$ for
    the moving window. (a). $x_{j} = 0$; (b). $x_{j} = 10$; (c).
    $x_{j} = 20$; (d). $x_{j} = 29$.}
  \label{fig:SzTimeMovingSmallW5}
\end{figure}

We now examine the local magnetizations at some specific
positions of the spin chain. In Fig.~\ref{fig:SzTimeMovingSmallW5}, we
plot $\langle S^{z}(x_{j},t)\rangle$ at $x_{j} = \{0, 10, 20, 29\}$.
We can see
that in all of four sub-plots $\langle S^{z}(x_{j},t)\rangle$ of moved
window fit quite well with the data from the case of of fixed window.
Especially, the further away from the initial perturbation point, the
longer the time scale for which we can obtain accurate results. This is easy to
understand as it takes longer time for the window to move to the
further sites of the chain, and the region to the left of the chain is
well-described by the Hilbert space of the semi-infinite chain. 
When the window passes through a region of the lattice,
these sites will eventually be absorbed into the right boundary, so this
causes the results measured after that time to be less accurate. That is
the reason why we see a big deviation of the results after some time.

\begin{figure}[ht]
  \centering
  \includegraphics[scale = 0.5]{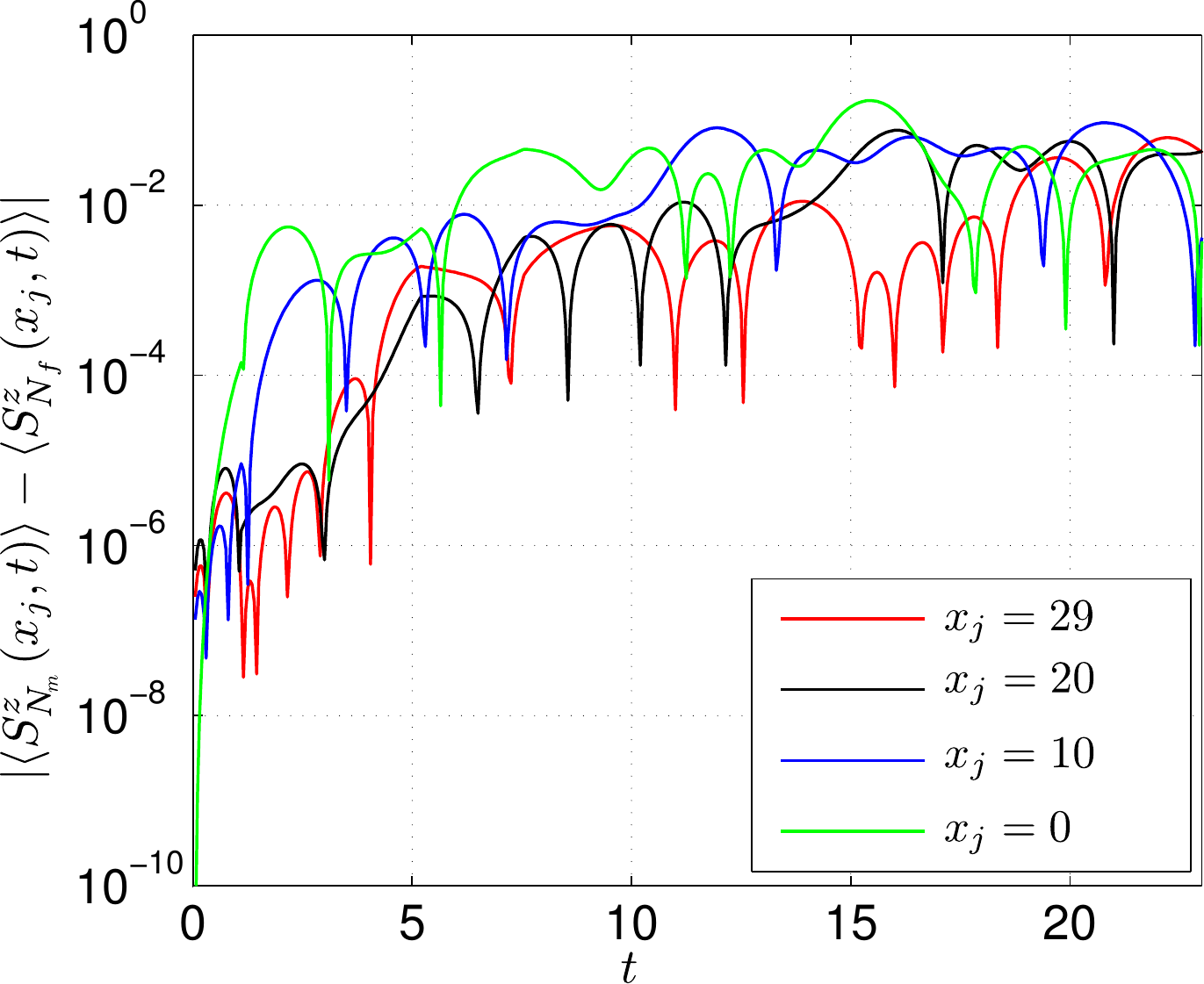}
  \caption{(Color online): Difference in local magnetizations evolving
    in time at specific positions $x_{j}$ on the left side of
    perturbation point between different schemes: fixed and moved
    window.}
  \label{fig:SzTimeMovingSmallW6}
\end{figure}

For a clearer comparison of $\langle S^{z}(x_{j},t)\rangle$, we also
plot the difference between two schemes in
Fig.~\ref{fig:SzTimeMovingSmallW6}. We can see that at the early time
the differences seem to be small and then increase in time. These
errors can be well-controlled by manipulating the size of the window
and the criteria for moving the window.

\section{Conclusion}
We have introduced two dynamical window techniques for studying the
real-time evolution of a locally perturbed infinite spin chain, the
expanding window and the moving window. Taking advantage of infinite
boundary conditions which has been introduced to replace an infinite
MPS by a finite MPS, we have proved that these two techniques 
are viable and are a more efficient replacement for the fixed window
technique.

One of the great advantages of these techniques a large saving in
computational resource as we only need to compute the evolution of a small
window of the system. This is most significant for the moving
window technique, where the computational cost per time-step doesn't
increase as the perturbation propagates through the system.
This is particularly relevant for cases where the physically relevant
dynamics of the system is a small region, for example in the vicinity
of the wave-front. For calculating quantities such as the spectral function, 
we are interested in regions only where the correlation function differs
significantly from zero, so this approach, where the dynamics are obtained
accurately in a small region and approximated elsewhere, is a significant
improvement.

\begin{acknowledgments}
  During preparation of this manuscript we learned of some related
  works\cite{TomotoshiMovingWindow,OsborneMovingWindow}.  We
  acknowledge support from the Australian Research Council Centre of
  Excellence for Engineered Quantum Systems and the Discovery Projects
  funding scheme (project number DP1092513).
\end{acknowledgments}

\footnotesize

\end{document}